\documentclass[conference]{IEEEtran}
\IEEEoverridecommandlockouts
\usepackage{cite}
\usepackage{amsmath,amssymb,amsfonts}
\usepackage{graphicx}
\usepackage{textcomp}
\usepackage{xcolor}
\usepackage{comment}

\usepackage{url}
\usepackage{subfig}
\usepackage{caption}
\usepackage{algorithm}
\usepackage[noend]{algorithmic}
\usepackage{booktabs} 
\usepackage{todonotes}

\usepackage{listings}
\newcommand{\subparagraph}{}
\usepackage{titlesec}
\titlespacing*{\section}{0pt}{3pt}{3pt}
\titlespacing*{\subsection}{0pt}{2pt}{2pt}
\usepackage{setspace}
\def\BibTeX{{\rm B\kern-.05em{\sc i\kern-.025em b}\kern-.08em
    T\kern-.1667em\lower.7ex\hbox{E}\kern-.125emX}}
\begin{document}

\title{URSA: Precise Capacity Planning and Contention-aware Scheduling for Public Clouds
}

\author{

\IEEEauthorblockN{1\textsuperscript{st} Ningxin Zheng, 2\textsuperscript{nd} Quan Chen}
\IEEEauthorblockA{
	\textit{Shanghai Jiao Tong University}\\
	Shanghai, China}

\and
\IEEEauthorblockN{3\textsuperscript{rd} Yong Yang}
\IEEEauthorblockA{
	\textit{Alibaba Cloud}\\
	Hangzhou, China }
\and
\IEEEauthorblockN{4\textsuperscript{th} Wei Zhang, 5\textsuperscript{th} Jin Li, 6\textsuperscript{th} Wenli Zheng, 7\textsuperscript{th} Minyi Guo}
\IEEEauthorblockA{
	\textit{Shanghai Jiao Tong University}\\
	Shanghai, China }

%

}

\maketitle

\begin{abstract}
Database platform-as-a-service (dbPaaS) is developing rapidly and a large number of databases have been migrated to run on the Clouds for the low cost and flexibility. Emerging Clouds rely on the tenants to provide the resource specification for their database workloads. However, they tend to over-estimate the resource requirement of their databases, resulting in the unnecessarily high cost and low Cloud utilization. A methodology that automatically suggests the ``just-enough'' resource specification that fulfills the performance requirement of every database workload is profitable. 

To this end, we propose \emph{URSA}, a capacity planning and workload scheduling system for dbPaaS Clouds. URSA is comprised of an online capacity planner, a performance interference estimator, and a contention-aware scheduling engine. The capacity planner identifies the most cost-efficient resource specification for a database workload to achieve the required performance online. The interference estimator quantifies the pressure on the shared resource and the sensitivity to the shared resource contention of each database workload. The scheduling engine schedules the workloads across Cloud nodes carefully to eliminate unfair performance interference between the co-located workloads. Our real system experimental results show that URSA reduces 25.9\% of CPU usage and 53.4\% of memory usage for database workloads while satisfying their performance requirements. Meanwhile, URSA reduces the performance unfairness between the co-located workloads by 47.6\% compared with the load balance strategy \emph{LeastRequestedPriority} in Kubernetes.
\end{abstract}


\section{Introduction}
\label{sec:intro}
Platform-as-a-service (PaaS) is a type of Cloud computing in which a service provider delivers a platform to tenants. Within the PaaS category, the fastest-growing segment is the database platform as a service (dbPaaS) according to Gartner's report~\cite{gartner}. The Cloud providers offer a variety of DBMS products, such as Amazon Relational Database Service (RDS)~\cite{awsrds} and Alibaba RDS~\cite{alibabards}. 
Most often, tenants rent powerful Cloud instances to ensure the good performance of their database workloads empirically. However, the empirical method often results in the excessive purchase of resources, invalidating the low price benefit of Cloud computing. Prior work~\cite{quasar} has shown that more than 90\% of Cloud applications apply for $5\times$ more resources than their actual demands. 

Capacity planning that identifies the smallest resource specification (number of cores, size of memory space) required by an application while its performance requirement can be satisfied is profitable for both tenants and Cloud providers. For tenants, renting Cloud instances of smaller specification reduces the cost. On the other hand, Cloud providers are able to serve more tenants with the same amount of hardware because the resource is better utilized.

To perform the capacity planning, a straightforward method is encouraging the tenants to profile their workloads under various resource specifications and find the ``just-enough'' resource specification that satisfies performance requirement. However, it is burdensome for tenants because it is time-consuming to collect the performance data on a large number of resource combinations. Moreover, whenever the workload changes, the tenant needs to profile the new workload. An alternative method is letting Cloud providers plan resource capacity for database workloads automatically. In this scenario, the goals are (1) identifying the ``just-enough'' resource specification for each database workload; (2) minimizing performance interference between workloads on the same nodes when scheduling the workloads. 

There are three challenges in achieving the above goals. First of all, Cloud providers have to identify appropriate resource specification for a database workload online quickly, because tenants would not provide their workloads for offline profiling due to privacy reasons. Second, only hardware event statistics and system-level indexes are available to plan the capacity. {\bf In other words, database-level statistics (such as the proportion of read and write operations, transactions per second) are also not available due to user privacy in the public Clouds}. Third, workloads run on the same node may contend for shared resources. Inappropriate workload scheduling may result in serious shared resource contention on some nodes while leaving the shared resource on other nodes free. In this case, the performance of the database workloads on the node under serious shared resource contention would be damaged. 

We propose URSA, a runtime system that consists of an \emph{online capacity planner}, a \emph{performance interference estimator}, and a \emph{contention-aware scheduling engine} to address the above challenges. URSA is proposed based on the observations that 1) hardware event statistics and system-level indexes strongly correlate with the performance of database workloads, and these statistics can be collected online by Cloud providers with negligible overhead; 2) for a database workload, if its performance is similar under two resource specifications, its ``pressure'' on shared resources and its sensitivity to the shared resource contention (i.e., performance degradation due to the contention) are also similar.

Specifically, the online capacity planner relies on a unified performance model that uses correlated hardware event statistics and system-level indexes as inputs. The training data is collected from the representative database workloads under multiple resource specifications. When a new database workload is received, the capacity planner runs it with the resource specification recommended by its owner, collects the hardware event and system-level statistics, predicts its performance under various resource specifications based on the corresponding pre-trained performance models, and identifies the ``just-enough'' resource specification for the workload to achieve the required performance. For each workload, the interference estimator quantifies its ``pressure'' on shared resources and its sensitivity to the shared resource contention. Based on the identified specification, the pressure on shared resources and the sensitivity to the contention of active workloads, the scheduling engine schedules the workloads carefully to avoid serious contention on shared resources.
The main contributions of this paper are as follows:

\begin{itemize}
	\item {\bf A precise online performance prediction method with negligible runtime overhead.} We show that system-level indexes and hardware event statistics strongly correlate with a workload's performance. Based on this observation, precise capacity planning for a workload without profiling it extensively offline is enabled. 
	\item {\bf A novel method to quantify performance interference due to shared resource contention for public Clouds.} URSA quantifies the pressure and sensitivity only based on system indexes and event statistics which means URSA is suitable for public Clouds.
	\item {\bf A novel contention-aware method to perform workload scheduling.} We show that prior workload scheduling methods may result in unfair performance slowdown of database workloads owing to the contention unawareness. We improve the performance fairness for database workloads in the Clouds.
\end{itemize}

Our experimental results show that URSA reduces 25.9\% of CPU usage and 53.4\% of memory usage for database workloads while satisfying their performance requirements. Meanwhile, URSA reduces the performance unfairness between the co-located workloads by 47.6\% compared with Kubernetes without affecting the system-wide throughput of our experimental dbPaaS Cloud.

\section{Related work}
\label{sec:related}
There has been some prior work on predicting the resource specification needed by a workload. 
Wu et.al~\cite{gpgpu} used clustering and classification techniques to predict the performance and power consumption for GPU. 
CloudScale~\cite{cloudscale} employs online resource demand prediction to achieve adaptive resource allocation. OtterTune~\cite{auto-parameter} presents an automated method to tune the software configuration knobs according to the past experience and machine learning skills. 
Quasar~\cite{quasar} measures the QoS of an application at a few resource configurations and uses collaborative filtering to predict the scale matrix of the target application. However, Quasar needs to measure the QoS status of the application that is invisible to public Cloud providers. When Quasar profiles an application, it needs to create a few copies of the target instance on other machines, which is costly and hurt user privacy. On the contrary, URSA predicts the scaling surface of a database only based on system-level indexes that are available online in public Clouds. 

Some other prior work focus on improving resource utilization in private data centers~\cite{bubble-up,prophet,bubble-flux,heracles,paragon,ubik} by co-locating the latency-sensitive applications with best-effort applications. In these datacenters, the operators understand the hosted applications and can perform offline profiling. Bubble-Up~\cite{bubble-up} quantifies the pressure and sensitivity of the application to the memory subsystem and judges whether the co-location is safe. 
Given a resource allocation for an unknown application, Paragon~\cite{paragon} predicts the impact of the interference on performance and assigns the application to an appropriate machine so that the applications do not interfere with each other. 
Heracles~\cite{heracles} dynamically manages multiple isolation mechanisms, such as the Cache Allocation Technology (CAT)\cite{intelrdt}, DVFS~\cite{dvfs}, to ensure the QoS of latency-sensitive applications. 
In general, prior work co-locates best-effort and latency-sensitive applications, and constraints the performance of best-effort applications based on the QoS status of latency-sensitive applications. However, in the public Cloud environment such as Amazon's RDS, every database workload is QoS sensitive. Besides, the QoS of database instance is invisible to the service provider. On the contrary, URSA quantifies the workloads' pressure and sensitivity to the shared resources contention based on system-level indexes. URSA can quantify the pressure and sensitivity without knowing the QoS of the target workload. 

\section{Motivation}
\label{sec:motivation}
In this section, we investigate the problems and challenges of capacity planning and workload scheduling in dbPaaS Clouds. 
In our investigation, we build a dbPaaS Cloud using 7 computer nodes connected with a 25Gb/s Ethernet switch. Table~\ref{tab:environment} shows the configurations of each node. To simulate real-system database workloads, we generate database workloads using two widely-used workload generators: Sysbench~\cite{sysbench}, and OLTP-Bench~\cite{oltpbench} that includes YCSB~\cite{ycsb}, TPC-C~\cite{tpcc}, LinkBench~\cite{linkbench} and SiBench~\cite{sibench} workloads. We adjust the configurations of Sysbench, YCSB, TPC-C, LinkBench, SiBench, and generate 11 variations for each of them. The $5\times 11=55$ workloads have different read-write ratios, compute densities, and database operating transactions, thus simulate a spectrum of real-system workloads. 

\newcommand{\tabincell}[2]{\begin{tabular}{@{}#1@{}}#2\end{tabular}} 
\begin{table}[hb]
	\centering
	\caption{\label{tab:environment}Hardware and software setup}
	\scriptsize
	\begin{tabular}{c|c}
		\hline
		& Configuration\\
		\hline
		Hardware & \tabincell{c}{CPU: Intel Xeon(R) Platinum 8163@2.50GHz\\Cores: 96; L3 shared cache: 32MB\\DRAM: 256GB; Disk: NVME SSD \\
			Network Interface Card (NIC): 25,000Mb/s} \\ 
		\hline
		Network & 25,000Mb/s Ethernet Switch \\
		\hline
		Software&\tabincell{c}{DBMS: AliSQL 5.6.32~\cite{alisql} \\ Operating system: Linux with kernel 3.10.0} \\
		\hline
	\end{tabular}
\end{table}

\begin{figure*}
	\centering
	\includegraphics[width=1.8\columnwidth]{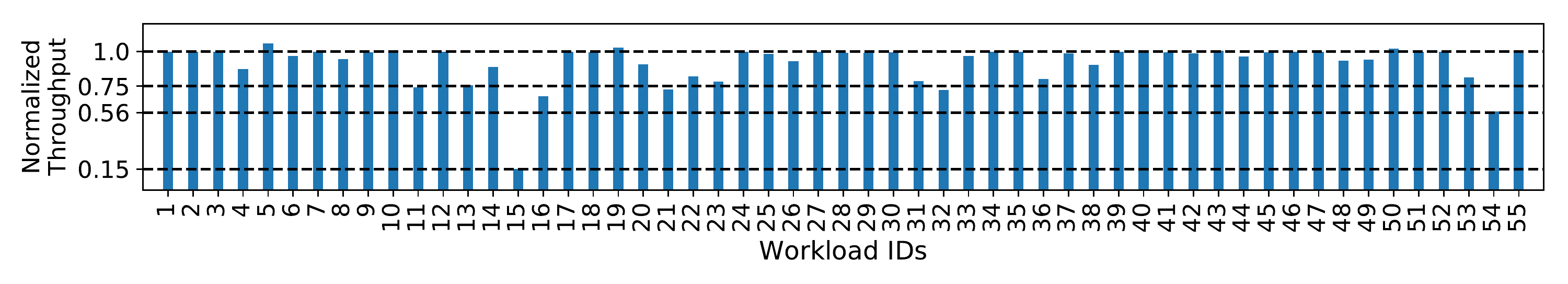}
	\vspace{-1mm}
	\caption{The normalized throughput of each workload when they are co-located on a dbPaaS Cloud.}
	\label{fig:moti slowdown}
	\vspace{-4mm}
\end{figure*}

\subsection{Existing Problems}
Without loss of generality, we assume that the largest Cloud instance for dbPaaS has 12 cores and 16 GB memory. 
We make this assumption because more than 70\% of the sold dbPaaS instances have less than 12 cores and 16 GB memory according to the statistics from our cooperative Cloud provider. The analysis is valid for both smaller and larger  instances.  
We search through all specifications for the smallest number of cores and size of memory required by each workload to achieve the same performance as the largest specification is used. The results show that 46.5\% of cores and 68.4\% of memory space are wasted on average if the tenants rent the largest Cloud instance. 
The resource waste may be worse if the largest instance has more cores and larger memory.

To identify the cost-efficient specification that satisfies performance requirement, a straightforward method is profiling the workload with all the specifications. However, the large search space results in the long profiling time. 
For instance, tenants can rent an instance that has up to 56 cores and 480GB memory in Alibaba RDS Cloud~\cite{alibabards}, thus there are up to $56\times 480=26880$ possible specifications if tenants are allowed to change the instance specification in the granularity of 1 core and 1GB memory. 
\begin{figure}
	\centering
	\includegraphics[width=.7\columnwidth]{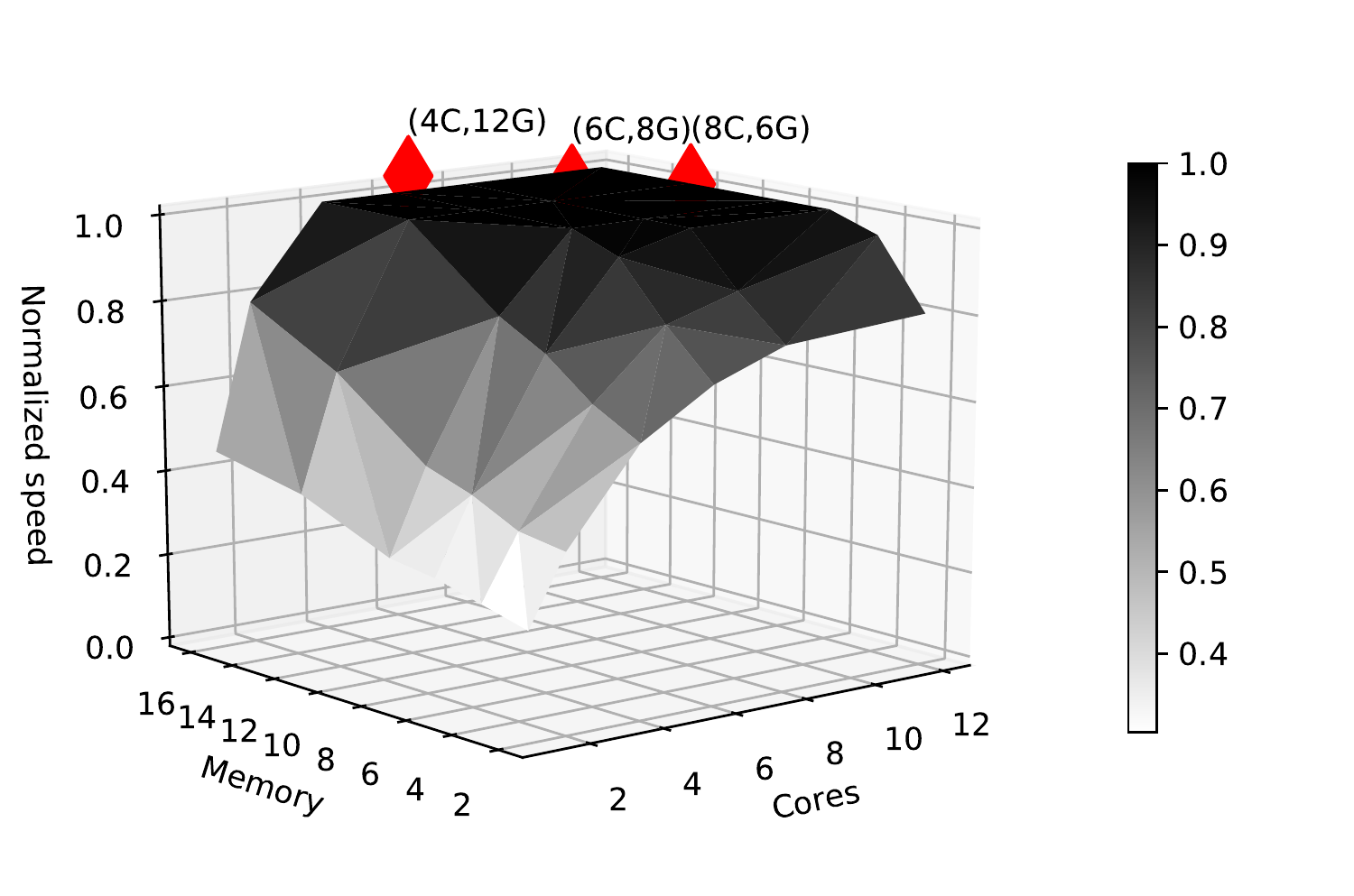}
	\caption{\label{fig:scaling}Scaling surface of workload $w$.}	
	\vspace{-4mm}
\end{figure}

Heuristic search is a solution to reduce the number of tries needed to identify the required specifications. However, it can get stuck in local optima. To show this problem, we collect the performance of a database workload $w$ under various resource specifications. Figure~\ref{fig:scaling} shows the {\it scaling surface} of workload $w$ that visualizes the performance of $w$ when its resource specification changes. In the figure, ($m$C, $n$G) represents the specification that has $m$ cores and $n$ GB memory. Observed from the figure, workload $w$ achieves the required performance with (4C, 12G), (6C, 8G), and (8C, 6G). In this scenario, the heuristic search may not identify the most cost-efficient specification. The heuristic search may stop at local optima, thus does not identify all the ``just-enough'' resource specifications. In this case, the identified specification may not be the most cost-efficient one considering the resource pricing policy. For instance, (8C, 6G) may be cheaper than (4C, 12G), but the search stops at (4C, 12G).

A more serious problem is that both exhausting search and heuristic search spend a long time on profiling the target workload. Depends on the characteristics of $w$, the number of tries needed to identify an appropriate specification could be large. For each try, we need to run workload $w$ and collect the performance data. Therefore, {\bf it is not practical to rely on either exhausting search or heuristic search to perform the capacity planning in the real system}. In contrast, URSA only executes the workload on a single specification and plan the capacity based on the predicted scaling surface directly.

To illustrate another problem, unfair sharing, we run the 55 database workloads with the largest specification and use {\it LeastRequestedPriority} policy~\cite{vohra2017scheduling} in Kubernetes~\cite{k8s} to schedule them in our experimental 7-node dbPaaS Cloud. {\it LeastRequestedPriority} is a wildly used load balancing policy in Kubernetes. 
The LeastRequestedPriority policy spreads out the resource consumption across the nodes~\cite{vohra2017scheduling}, thus it avoids serious performance interference between workloads caused by the unbalanced resource distribution. 

Figure~\ref{fig:moti slowdown} shows the throughput of each workload in the Cloud when they are scheduled by the LeastRequestedPriority policy normalized to its solo-run throughput. Observed from the figure, some workloads are slowed down much worse than the others (e.g., workload-15). If workloads that stress on the same shared resource are scheduled to the same server, their performance is seriously damaged. 
The unfair performance degradation is due to the unawareness of the ``pressure'' of each workload on shared resources and the sensitivity of each workload to the shared resource contention. 
{\bf Workloads suffer from unfair performance degradation in Clouds even if the cores and memory space are fairly allocated on every node.} 

\subsection{Guidelines of URSA}
To resolve the above problems, we propose URSA, a runtime system that identifies the most cost-efficient specification for each database workload while its performance requirement is satisfied, and schedules workloads to eliminate the unfair sharing. We design URSA following four guidelines. 
\begin{itemize}
	\item To identify the appropriate specification for a database workload online, URSA should be able to obtain the scaling surface (Figure~\ref{fig:scaling}) of the workload without profiling it with a large number of specifications.
	\item URSA should be able to obtain the pressure on the shared resource and the sensitivity to the shared resource contention of each database workload.
	\item URSA should achieve the above two goals based on system-level statistics which is the only available information in public Clouds, because tenants may not provide application-level information due to privacy. 
	\item URSA should be able to balance database workloads while avoiding serious contention on shared resources. 
\end{itemize}


\section{Design of URSA}
\label{sec:design}
\begin{figure}
	\centering
	\includegraphics[width=.8\columnwidth]{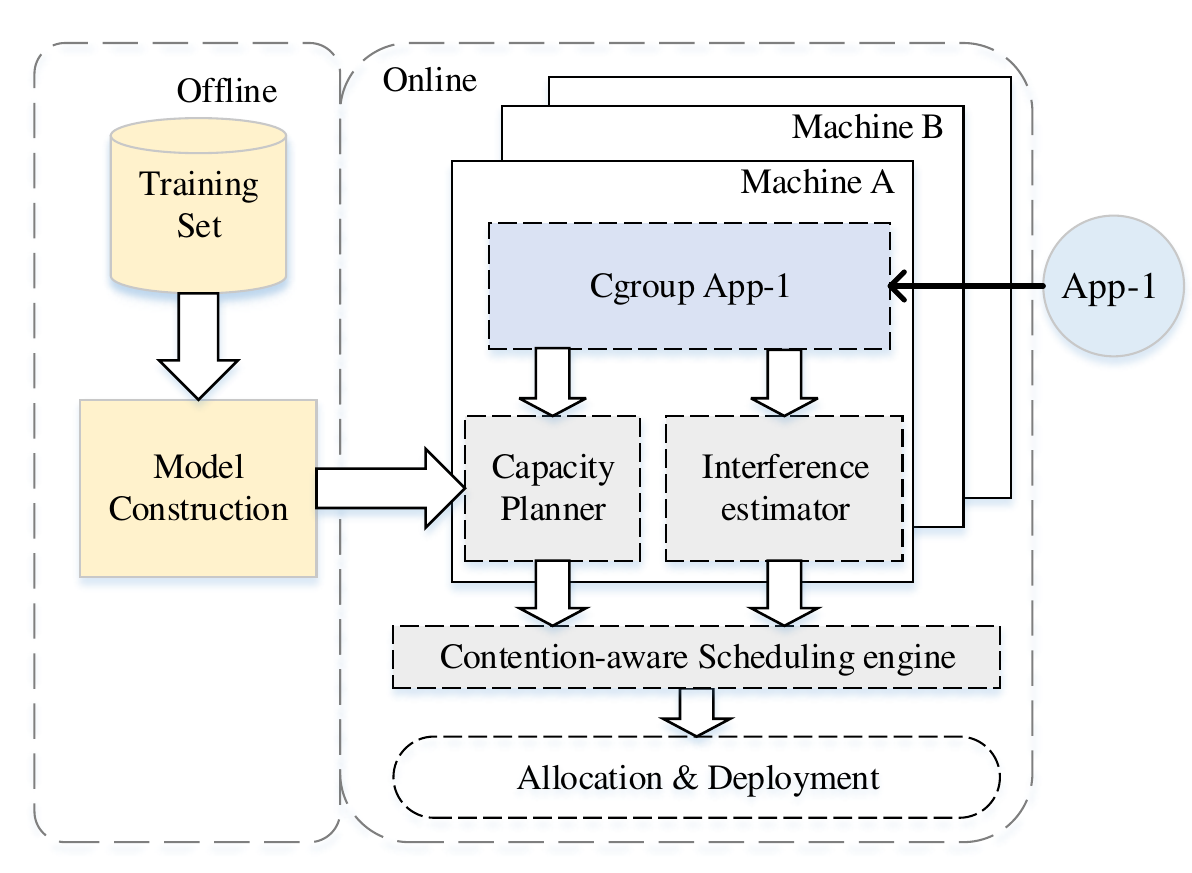}
	\caption{\label{fig:overview}Design overview of URSA.}
	\vspace{-3mm}
\end{figure}

Figure~\ref{fig:overview} shows the design overview of URSA. URSA deploys an {\it online capacity planner} and a {\it performance interference estimator} on each node, and a {\it contention-aware scheduling engine} at the Cloud level. 
The capacity planner precisely predicts the scaling surface of a workload and plans the appropriate resource specification for the workload based on the scaling surface and the performance requirements. The interference estimator quantifies the pressure of a workload on the shared resources and its sensitivity to the shared resource contention. The scheduling engine distributes the database workloads across the Clouds so that they would not suffer from serious slowdown due to shared resource contention.

In more detail, URSA performs capacity planning for a new database workload $w$ in four steps. 1) URSA migrates $w$ to run alone under the specification recommended by its owner for a short time and collects performance counter statistics and system-level indexes required by the capacity planner. 2) Based on the collected information and a unified performance model, the capacity planner predicts $w$'s scaling surface as shown in Figure~\ref{fig:scaling}, and identifies the appropriate specification for $w$ to fulfill its performance requirement. 3) The interference estimator launches a set of carefully designed micro-benchmarks that pressurize the shared resources respectively, such as the LLC and memory bandwidth, etc, on the node that runs $w$. By adjusting the pressures of the micro-benchmarks on the shared resources, monitoring the system-level indexes of $w$, the interference estimator quantifies $w$'s pressure on various shared resources and its sensitivity to the contention on shared resources under the specification recommended by URSA. 4) Based on the pressure and sensitivity of $w$, the scheduling engine calculates the possibility of $w$ interfering with the already deployed workloads on each node. Based on the possibility, $w$ is assigned to the node with enough hardware resources yet has the smallest interference possibility.

The capacity planner can implement multiple planning policies to fulfill different design purposes. For instance, it can be implemented to identify the ``just-enough'' specification that does not degrade $w$'s performance compare with the original specification provided by the user. It can also be implemented to recommend a cost-efficient specification for $w$ to satisfy a specific QoS target. Our evaluation in Section~\ref{sec:Case} shows that both the two policies are effective and efficient.

It is possible that the workload intensity and resource demand change during the execution, invalidating the current ``just-enough'' specification, in real-world systems. In this scenario, URSA is able to accommodate the workload characteristic to provide consistently satisfactory performance. Once the performance model is built, URSA only needs to collect the system-level performance statistics to predict the performance scaling surface of the workload. {\bf When request spike happens, URSA can dynamically collect the statistics, predict its current scaling surface with negligible overhead, and adjust the specification accordingly}.

Although URSA is proposed for real system dbPaaS Clouds that only host databases (e.g., Alibaba RDS), {\bf it can be generalized for Clouds that host general applications as long as more training samples are collected from them to capture their scaling characteristics}. 






\section{Online Capacity Planning}
\label{sec:SSP}
While planning capacity, the capacity planner first predicts the scaling surface of the target workload based on the hardware events and system-level indexes. Then, the capacity planner uses different planning policies to adjust the resource specification according to scenarios and requirements. 

The insight of the capacity planner is that some system-level indexes and performance event statistics of a workload, such as instruction per cycle (IPC), can reflect its scale characteristic. IPC can be used to determine whether a workload is CPU-bound or memory-bound. The performance of the CPU-bound workload is positively related to the number of allocated cores. Therefore, if the indexes and performance event statistics of two workloads are close, then these two workloads tend to have similar scale characteristics. If a workload is far different from the workloads in the training set, URSA can periodically update the model using the incremental update. Moreover, the service provider has a wide variety of workloads to generate good coverage of the scaling characteristic space.
\subsection{Identifying Correlated Features}
The operating system provides low overhead tools to collect the performance statistics of the application, such as perf~\cite{perf} and cgroups~\cite{cgroups}. 
A brute force method is incorporating all available system-level indexes in the capacity planner. However, too many irrelevant indexes may lead to overfitting and low accuracy and also increase the training and prediction time of the capacity planner.
Thus, we use Lasso regression~\cite{lasso} to identify the features related to the workload's performance. Accurately, URSA first collects various hardware counter events and system-level indexes from the representative workloads. Then, URSA uses Lasso to filter out the most relevant system-level indexes and hardware counters of the workload's performance. The filtered system-level indexes and hardware counter events are shown in Table~\ref{tab:indexes}.

\begin{table}[thb]
	\centering
	\vspace{1mm}
	\caption{\label{tab:indexes}System-level indexes and hardware counter events}
	\scriptsize
	\begin{tabular}{l|l||l|l}
		\hline
		Index & Source& Index &Source\\
		\hline
		(1) IPC & perf& (9) page-fault&perf \\
		(2) dTLB-store-misses&perf&(10) dTLB-load-misses&perf\\
		(3) cache-misses&perf&(11) cache-references&perf\\
		(4) node-stores&perf&(12) node-loads&perf\\
		(5) io-read-bytes&cgroup&(13) io-write-bytes&cgroup\\
		(6) io-serviced-read&cgroup&(14) io-serviced-write&cgroup\\
		(7) memory usage&cgroup&(15) dirty memory&cgroup\\
		(8) cpu usage&cgroup & & \\
		
		\hline
	\end{tabular}
	\vspace{-3mm}
\end{table}
\subsection{Capacity Planner Construction}
\label{sec:ssp training}
\begin{figure}
	\centering
	\includegraphics[width=.8\columnwidth]{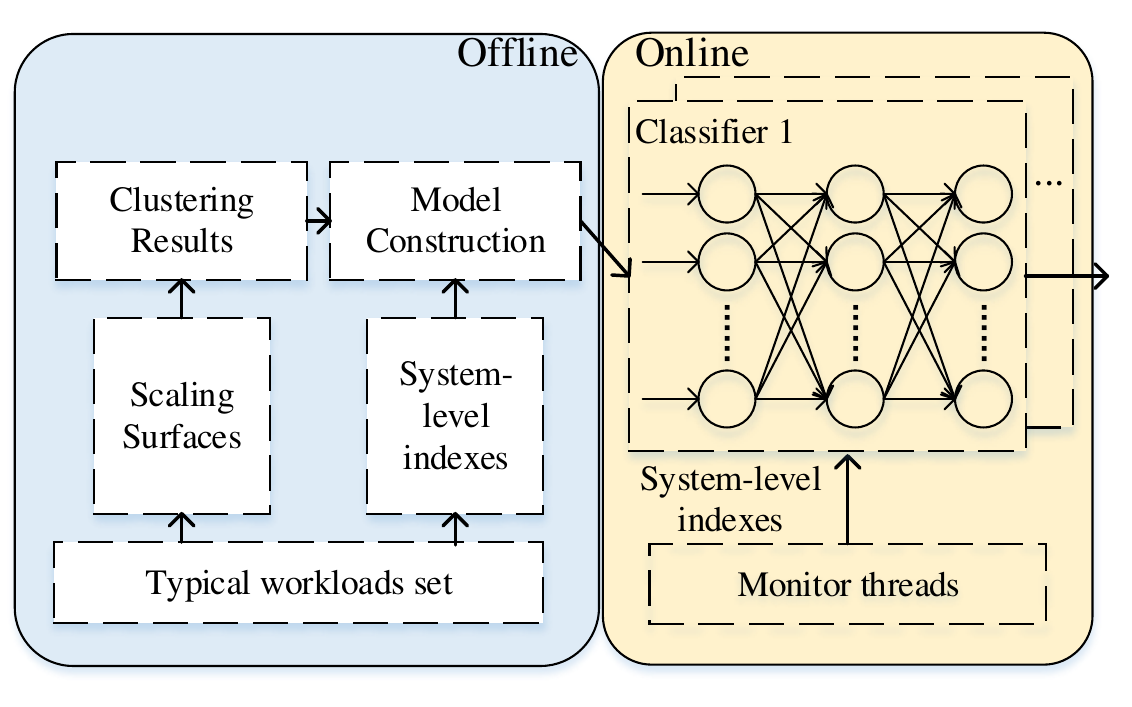}
	\caption{Design of the capacity planner.}	
	\label{fig:ssp-design}
	\vspace{-4mm}
\end{figure}


Figure~\ref{fig:ssp-design} shows the design of the capacity planner. First, the capacity planner collects the scaling surfaces and the selected indexes of a set of representative workloads in a specified configuration region offline. Second, the capacity planner clusters a set of scaling patterns according to the collected scaling surfaces. To build the training set of the online classifier, URSA aggregates the system indexes and event statistics of the applications that have the same $ClusterID$. Specifically, the training set is built as $\langle System Indexes, ClusterID\rangle$ where $ClusterID$ is the clustering result of the application's scaling surface. $System Indexes$ is the average values of the performance statistics of the workload for each time period (such as 10 seconds) during the execution. Each cluster's center scaling surface is the mean vector of the other scaling surfaces. The center scaling surface is used as the representative scaling surface of this cluster for specification searching, which is reasonable because the scaling surfaces of workloads in the same cluster are similar. Finally, the capacity planner uses $ System Indexes$ as input, $ClusterID$ as output to train the classification model.

Every capacity planner has one or more base configurations, and each base configuration corresponds to a classifier. The $System Indexes$ used to train the classifier are collected under the corresponding base configuration. For the scenario in which the workloads' resource configuration cannot be changed during the capacity planning, the capacity planner can train a classifier for each configuration in the configuration region. Then, the capacity planner selects the current configuration of the target workload as the base configuration and uses the corresponding classifier to predict its scaling surface. In this case, the capacity planner can perform capacity planning without customers' perception. It is worth noting that training a classifier for each base configuration will not introduce more data collection overhead, because the \emph{System Indexes} can be collected at the same time while collecting the scaling surface. 

When URSA performs capacity planning for an online workload, the monitor thread collects the system-level indexes and hardware counter statistics of the workload under the base configuration and passes them to the trained classifier. The classifier classifies the workload into a scaling surface cluster according to the collected indexes. The capacity planner returns the representative scaling surface of the cluster as the predicted scaling surface of the target workload. 

According to different scenarios, the capacity planner uses different planning policies to adjust the resource configuration. In the scenario of specification extension, the capacity planner searches for the minimum specification in the predicted scaling surface that can meet the users' speedup requirements. In the scenario of improving utilization and cost performance, the capacity planner recommends a smaller resource configuration that has the same performance with the current specification.


\subsection{Prediction Accuracy}
\label{sec:prediction accuracy}

We evaluate the capacity planner using 55 workloads (mentioned in Section~\ref{sec:motivation}) in the configuration region: [1 core-12 cores, 2GB-16GB]. The workloads set is randomly divided into a training set containing 44 workloads and a validation set containing 11 workloads. We use the transaction per second (TPS) as the performance indicator of the workloads and train the capacity planner as mentioned in section~\ref{sec:ssp training}. Specifically, the K-means is used to cluster the scaling surfaces and the Multilayer Perceptron\cite{mlp} is used as the classifier in the capacity planner. Equation~\ref{eq:surface-dis} defines the error between the predicted scaling surface and the actual scaling surface of the target workload. In this equation, $N_{conf}$ is the number of resource configuration in the configuration region, $Speedup_{i}$ is the predicted speedup of the configuration~\emph{i} relative to the base configuration, and the $Speedup'_{i}$ represents the actual speedup of the configuration~\emph{i} relative to the base configuration.
\begin{equation}
\label{eq:surface-dis}
Err_{surface}=\sum_{i=1}^{N_{conf}} |\frac{Speedup_{i}}{Speedup'_{i}}-1| /N_{conf}
\end{equation}
\vspace{-2mm}

\begin{figure}
	\centering
	\includegraphics[width=0.7\columnwidth]{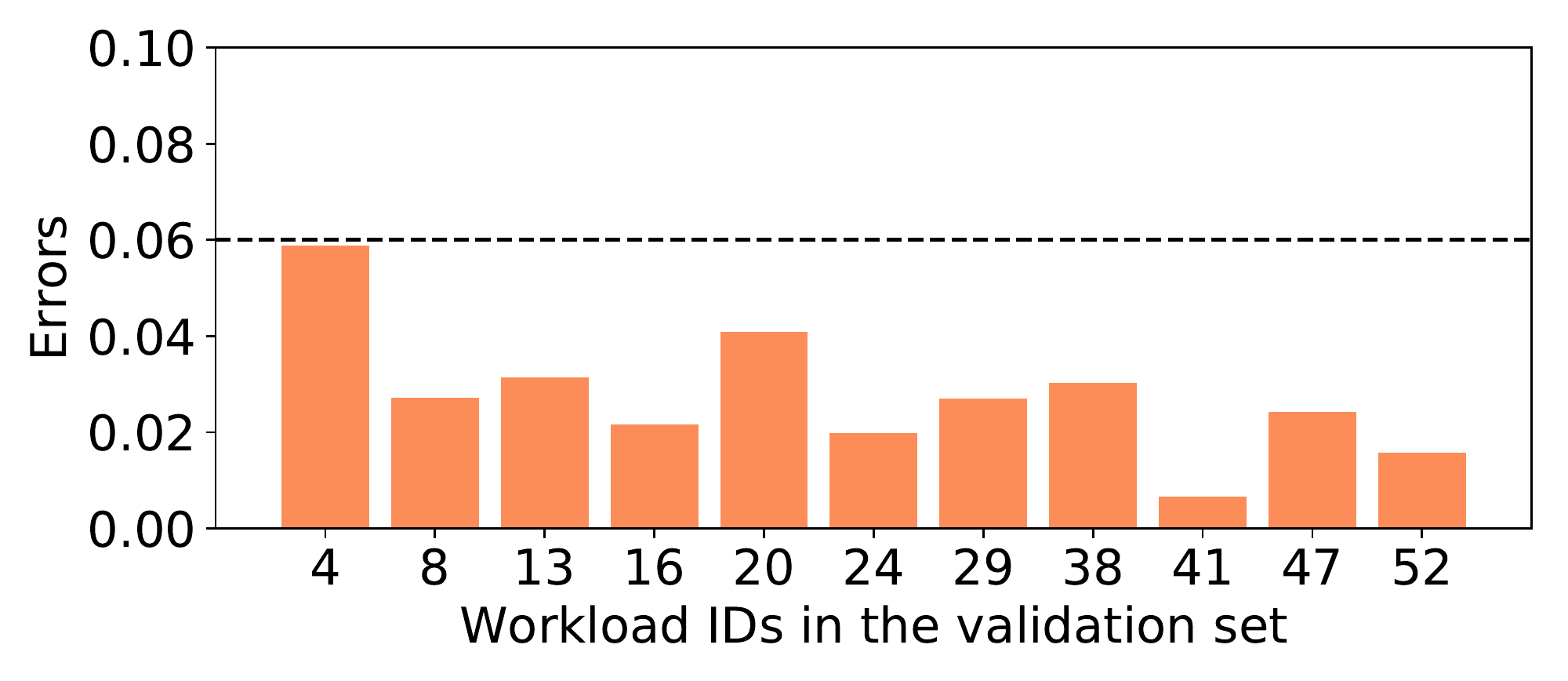}
	\caption{Prediction errors of the capacity planner.}
	\label{fig:accuracy}
	\vspace{-2mm}
\end{figure}

Figure~\ref{fig:accuracy} shows the prediction errors of the workload in validation set when the \emph{(6C, 8G)} is selected as the base configuration and the number of clusters in K-means is 20. Observed from Figure~\ref{fig:accuracy}, the maximum prediction error of validation workloads equals to 5.88\%. Thus, the capacity planner can accurately predict the scaling surface of the workload based on the hardware counter events and the system-level indexes. 

\begin{figure}[htbp]
	\centering
	\includegraphics[width=0.75\columnwidth]{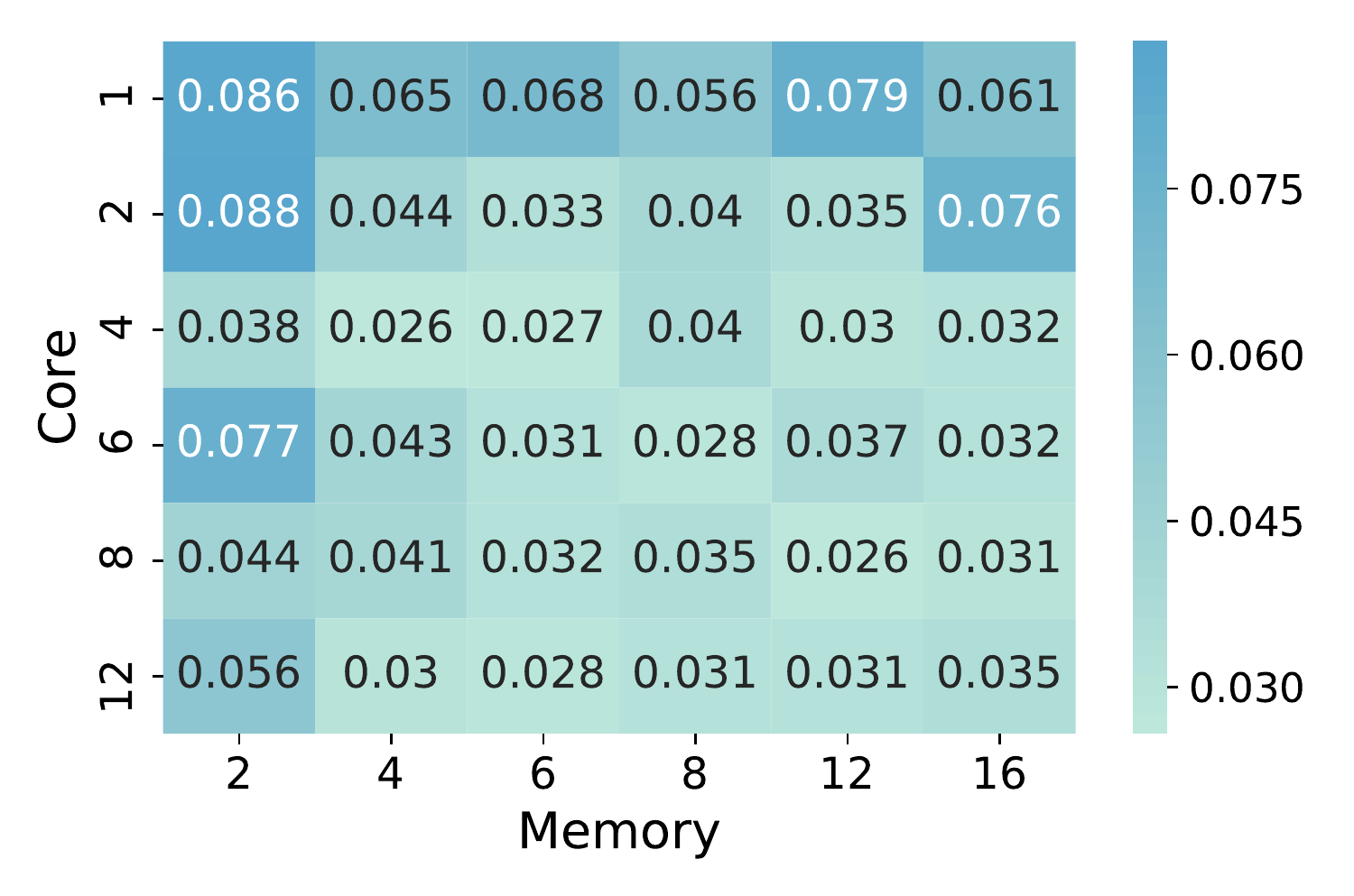}
	\vspace{-1mm}
	\caption{Prediction errors with different base configs.}
	\label{fig:base selection}
	\vspace{-3mm}
\end{figure}

\subsection{Sensitivity to Hyper Parameters}
\label{sec:hyper para}

This experiment shows how to select the base configuration and the number of the clusters. Figure~\ref{fig:base selection} shows the average prediction errors of the validation workloads (shown in Figure~\ref{fig:accuracy}) when selecting different base configurations. As shown in Figure~\ref{fig:base selection}, configurations in the middle of the region have a lower prediction error and the error rises when a certain dimension of the resource configuration is pretty low. Therefore, in the scenario that the capacity planner can change the workloads' resource configuration during the capacity planning, it is better to choose the configurations in the middle of the region as the base configuration. If the workloads' resource configuration cannot be changed during the capacity planning, the capacity planner can train a classifier for each configuration in the configuration region, and use the corresponding one to predict the scaling surface for the target workload.

\begin{figure}
	\centering
	\includegraphics[width=0.75\columnwidth]{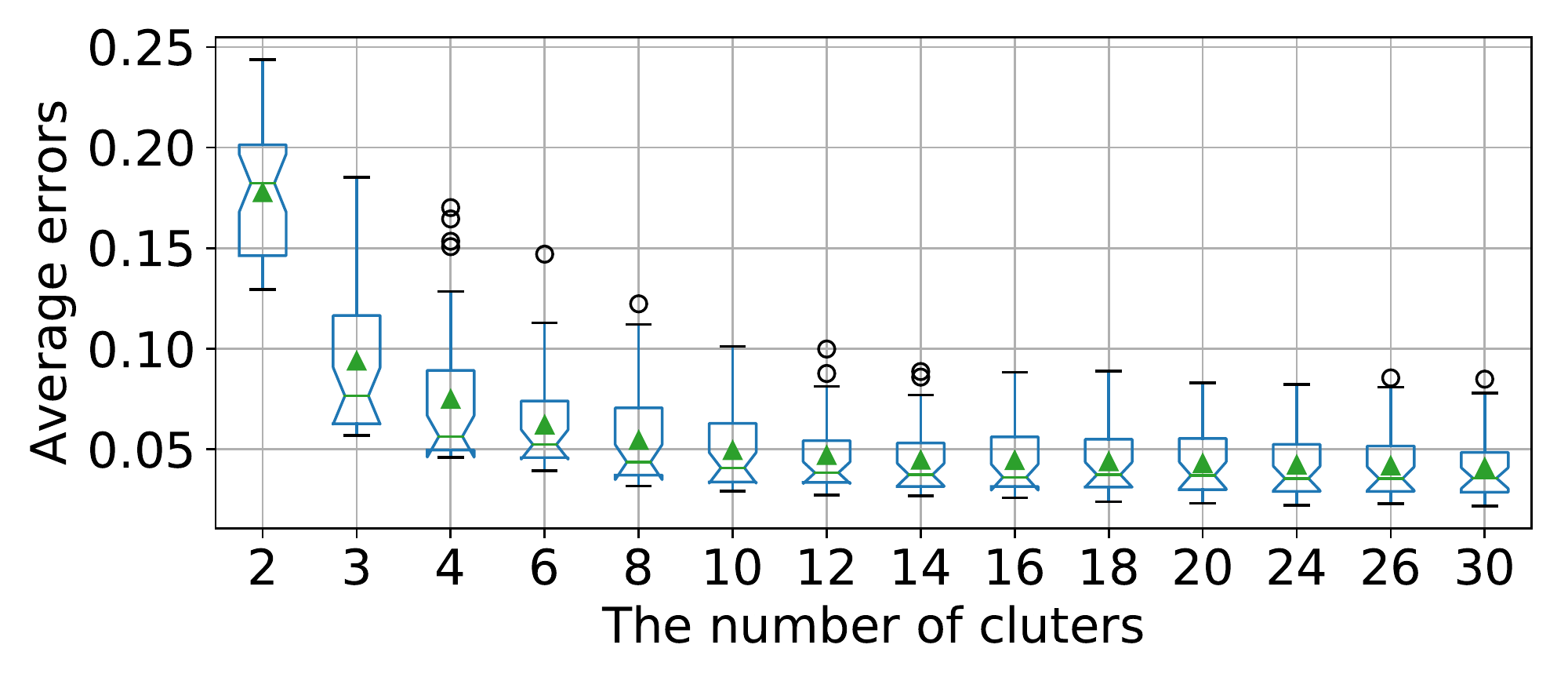}
	\caption{Average errors of different cluster numbers.}
	\label{fig:cluster selection}
\end{figure}

The number of clusters $K$ is a parameter of the K-means algorithm. On one hand, too few clusters may force dissimilar scaling surfaces into one cluster and lead to low accuracy of the classifier. On the other hand, too many clusters may lead to overfitting and high complexity of the model. To evaluate the sensitivity of prediction error to the number of the clusters in K-means, we measure the average prediction error of the validation set mentioned above as we vary the number of the clusters, as shown in Figure~\ref{fig:cluster selection}. Specifically, we use 2, 3, 4,\dots, 30 as \emph{K} in K-means to train a total of 432 capacity planner models for 36 configurations in the configuration region. In Figure~\ref{fig:cluster selection}, the y-axis represents the average error of trained capacity planner models on the validation set. We can see that the prediction error decreases with the increase of the number of clusters and is basically unchanged after a certain size. Considering both the training time and accuracy, we select 20 as the number of clusters.

\begin{figure*}[htbp]
	\centering
	\includegraphics[width=1.85\columnwidth]{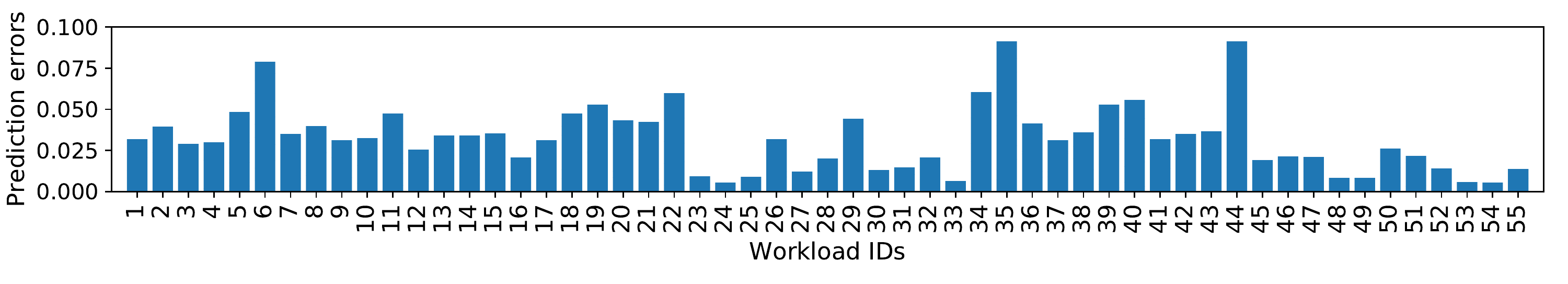}
	\vspace{-2mm}
	\caption{Prediction errors of the capacity planner with the leave-one-out cross validation.}
	\label{fig:loocv}
	\vspace{-3mm}
\end{figure*}

Besides, we also use leave-one-out cross validation to evaluate the accuracy of the capacity planner. In each round, an application is selected as the test set and the other workloads are used as the training set. The number of the cluster is 20 and the base configuration is \emph{(6C, 8G)}. The maximum prediction error of all 55 rounds is 9.1\% and the average prediction error of all rounds is 3.23\% (as shown in Figure~\ref{fig:loocv}). So, the capacity planner can accurately predict the scaling surfaces of the workloads.

\section{Interference Estimating}
\label{sec:PSQ}
To alleviate the interference between the target workload and the other co-located workloads, URSA also quantifies the pressure of the target workload on the shared resources and its sensitivity to the contention on shared resources. Specifically, the interference estimator gradually increases the pressure on different shared resources respectively until the corresponding system index reaches the threshold. Then, the interference estimator quantifies the pressure and the sensitivity to different shared resources based on the collected system-level indexes. 

\subsection{Interference due to LLC}
The interference estimator uses kilo LLC misses per second (kmps) to quantify the pressure on LLC. When a cache miss happens, the operating system loads data into a cache line and replaces the old data. So if we do not isolate the LLC and the location of the cache access is random, then from the perspective of probability, the program that triggers more LLC misses will occupy more cache lines in the competition of the LLC. In other words, workloads with higher kmps will produce more pressure on the LLC. Therefore, the interference estimator uses kmps to quantify the pressure on LLC produced by the target workload. 

We design some LLC stress programs with different pressure levels. 
The stress programs randomly access to buffers of different sizes. Accessing larger buffers will generate more cache misses and more pressure to the LLC. The interference estimator uses CAT to collect the kmps values of different levels stress programs in different LLC sizes offline. When quantifying the pressure and sensitivity to LLC, URSA uses CAT to gradually narrow down the LLC size allocated to the target workload until the kmps value rises by 10\%. The interference estimator records the target workload's kmps track and compares it with different levels of stress program. URSA uses CAT to pressurize the LLC rather than the stress programs mentioned above, because the stress programs will pressurize the LLC and memory bandwidth at the same time. The distance between the kmps tracks of two workloads is defined as $\sum_{w=1}^{W} (kmps_w-kmps'_w)^2$, where \emph{W} is the number of the cache ways in LLC, $kmps_w$ and $kmps'_w$ are the kmps values of the target workload and the stress program respectively. 
The interference estimator uses the closest stress program's pressure level as the pressure level of the target workload. The sensitivity of the target workload is quantified as the number of cache ways when the kmps rises by 10\%.

\subsection{Interference due to Memory Bandwidth}
The interference estimator uses the memory bandwidth usage to quantify the pressure and sensitivity to the memory bandwidth. Equation~\ref{eq:mbw pressure} shows the quantification of pressure to the memory bandwidth, in which $N_{mbw}$ is a constant and represents the number of pressure levels, $Usage_{mbw}$ is the memory bandwidth usage of the target workload, and $Phy_{mbw}$ represents the physical memory bandwidth of the machine. 
\begin{equation}
\vspace{-1mm}
\label{eq:mbw pressure}
Pressure_{mbw}=N_{mbw}\times\frac{Usage_{mbw}}{Phy_{mbw}}
\end{equation}

To quantify the sensitivity to memory bandwidth, the interference estimator also designs some stress program for memory bandwidth. The memory bandwidth stress programs access a buffer continuously in steps of the size of a LLC cache-line. In this case, each memory access of the stress program triggers a LLC cache miss and consumes the memory bandwidth to load the data into LLC. The interference estimator generates some stress programs with different pressure levels by adjusting the size of the buffer and the number of threads of the stress program. When quantifying the sensitivity, the interference estimator gradually increases the pressure level of the stress program until the target workload's memory bandwidth usage drops by 10\%. The workload's sensitivity to the memory bandwidth contention is quantified as $N_{mbw}-Max_{mem}$, where $N_{mbw}$ is the number of pressure levels and $Max_{mem}$ is the maximum pressure level of the stress program that the workload can withstand. 

However, the stress programs consume not only memory bandwidth but also LLC. Therefore, when using the stress programs to pressurize the memory bandwidth, it is difficult to determine whether the decrease of the workload's memory bandwidth usage is caused by the LLC contention or the memory bandwidth contention. So, the interference estimator also uses CAT to limit the stress program to use the least LLC. In this case, the quantification of the sensitivity to memory bandwidth will hardly be affected by the LLC contention.

\subsection{Interference due to Disk}
Compared to the disk bandwidth, DBMS is more sensitive to the I/O operations per second. Therefore, the interference estimator uses I/O operations per second (IOPS) to quantify the pressure and sensitivity to the disk for DBMS. 
The pressure to disk is quantified as $IOPS/IOPS_{Scaler}$, where $IOPS$ is the IOPS of target workload and $IOPS_{Scaler}$ is a constant representing the iops of the unit pressure level. 

To quantify the workload's sensitivity to the disk contention, the interference estimator uses fio\cite{fio} to generate some stress programs with different pressure levels. Specifically, the interference estimator adjusts the depth of the I/O queue and the number of the threads in the fio to produce different levels of pressure to the disk. The interference estimator gradually increases the pressure level of the stress program until the IOPS of the target workload drops by 10\%. The sensitivity of the workload to the contention on disk is quantified as $N_{disk}-Max_{disk}$ where $N_{disk}$ is the number of pressure levels of the stress programs and $Max_{disk}$ is the maximum pressure level that the target workload can withstand.

\subsection{Interference due to Network}
The interference estimator uses the network bandwidth usage to quantify the pressure to the network produced by the target workload. The pressure to network is quantified as \mbox{$N_{nbw}\times Usage_{nbw}/Phy_{nbw}$}, where $N_{nbw}$ represents the number of the pressure levels, $Usage_{nbw}$ is the network bandwidth usage of the target workload, and $Phy_{nbw}$ is the physical network bandwidth. In addition, the interference estimator uses iperf3\cite{iperf} as the stress program to quantify the sensitivity to the network. Specifically, the interference estimator increases the network bandwidth consumed by the stress programs gradually until the network bandwidth usage of the target load drops by 10\%. The sensitivity to the contention on network is quantified as $N_{nbw}-Max_{nbw}$ where $N_{nbw}$ represents the number of the pressure levels of the stress programs and $Max_{nbw}$ is the the maximum pressure level of the network contention that the target workload can withstand.


\section{Contention-Aware Scheduling}
\label{sec:APE}
In this section, we present the details of the contention-aware scheduling engine. The engine selects an appropriate machine to deploy the target workload based on the quantified pressure and sensitivity on the shared resources.
In the public cloud environment, the customers' instances are equally important and every instance is QoS sensitive. Therefore, the service provider should ensure fairness when deploying workloads. Even if resource isolation technology can isolate CPU and memory, the workloads may still interfere with each other on the shared resources. Unmanaged interference may result in unfairness in the public cloud. Specifically, the competitive pressure imbalance of different shared resources on the same machine will lead to certain shared resources becoming bottlenecks. In addition, the unbalanced competition of shared resources between different machines also seriously damage the fairness. 

Equation~\ref{eq:contention risk} defines the contention risk on a machine, where $N_R$ is the number of the shared resources, $MaxS_r$ represents the maximum sensitivity level of all deployed instances to this shared resource, $SumP_r$ is the sum of the pressure levels of the deployed instances, and $SCALER$ is a constant greater than one, such as 1.1. $SumP_r$ reflects the pressure to the shared resource produced by all the deployed workloads. $MaxS_r$ reflects the pressure on this shared resource that the most sensitive deployed instance can withstand. The larger the $MaxS_r$, the less pressure the most sensitive instance can withstand. We also use $SCALER$ to balance the contention intensity between different shared resources of the same node. As the sum of pressure levels of a certain shared resource increases, the contention risk of this resource also grows exponentially. Therefore, in order to reduce the risk of interference, the scheduling engine avoids excessive pressure of a certain shared resource when deploying instances.
\vspace{-1.5mm}
\begin{equation}
\label{eq:contention risk}
Pro_{c}=\sum_{r=1}^{N_R}(MaxS_{r}\times SumP_{r}\times SCALER^{SumP_{r}})
\end{equation}
\vspace{-1mm}
$Pro_{c}$ reflects the possibility of workloads interfering with each other. Besides, the scheduling engine also considers the balance of consumed resources between nodes. When a new instance arrives, the scheduling engine scans and scores the nodes that have enough resource to deploy this instance. The score of each node is calculated in Equation~\ref{eq:score} in which $Usage_{Ave}$ is the average percentage of the used CPUs and memory of the node.  The scheduling engine will deploy the coming instance to the machine with the smallest $Score$. Therefore, the scheduling engine tends to deploy the instance to machines with less contention risk and more available resources.
\begin{equation}
\label{eq:score}
Score=Pro_{c}\times Usage_{Ave}
\end{equation}

The contention-aware scheduling engine considers the contention of the shared resources and the resource distribution at the same time. Therefore, 
The scheduling engine can avoid the serious unfairness caused by the unbalanced resource distribution and the contention on the shared resources. 
The main advantages of the contention-aware scheduling engine are: (1) The scheduling engine can balance the contention risk between different nodes. (2) The scheduling engine also balances the contention intensity between the shared resources of the same node. (3) The scheduling engine also takes the resource distribution into account which avoids the unfairness caused by the unbalanced resource allocation.

\section{Evaluation of URSA}
\label{sec:Case}
In this section, we evaluate the accuracy of the capacity planning in URSA, followed by the effectiveness of the contention-ware scheduling engine in alleviating shared resource contention. 
\begin{figure*}[tbp]
	\centering
	\includegraphics[width=1.75\columnwidth]{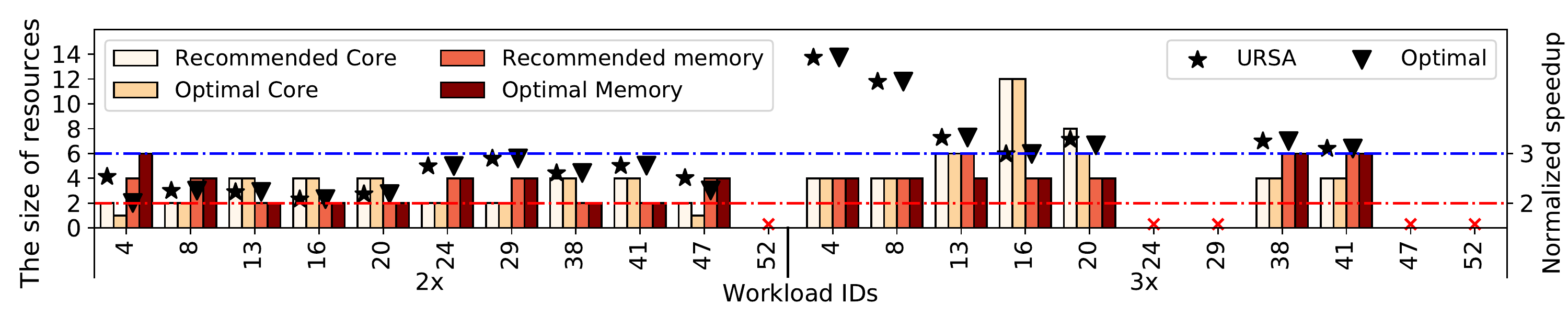}
	\vspace{-1mm}
	\caption{Specifications recommended by URSA and the optimal specifications for achieving performance target.}
	\label{fig:solo_up}
	\vspace{-3mm}
\end{figure*}

When evaluating the accuracy of the capacity planning, we consider two existing scenarios. In the first scenario, a tenant finds its current specification does not satisfy the performance requirement and requests a higher but cost-efficient specification. In this case, the tenant needs to provide the target performance normalized to the performance with the current specification. 
In the second scenario, a tenant finds the performance of its workload is satisfactory and requests a lower specification that does not hurt the current performance to reduce the cost. The performance of database workloads is defined as {\it transactions per second (TPS)} in this section.

\subsection{Scenario 1: Achieving Performance Target}
\label{sec:case 1}
In this experiment, we use the same workload set and experiment environment as described in Section~\ref{sec:prediction accuracy}. From the 55 workloads, we randomly choose 44 of them to train the performance model offline and use the rest 11 workloads as the validation set. When train the performance model offline, according to our investigation in Section~\ref{sec:prediction accuracy}, we set the number of clusters in the clustering algorithm to be 20. For each workload in the validation set, we assume its origin resource specification is (1C, 2G) that consists of 1 core and 2GB memory. URSA is allowed to recommend a new resource specification ($m$C, $n$G) for a workload, where $m$ and $n$ are randomly integers ($1\le m \le 12$, $2 \le n \le 16$). 

Figure~\ref{fig:solo_up} shows the specification recommended by URSA if the expected performance is 2 times and 3 times of the performance with the current resource specification. To show the effectiveness of the capacity planning,  Figure~\ref{fig:solo_up} also shows the optimal (i.e., the smallest) specification identified by searching through all the possible resource specifications. In the figure, $\times$ means that the required performance improvement cannot be satisfied even with the largest resource specification. 

Observed from the figure, URSA recommends the optimal resource specifications for 18 out of the 22 requests. For the other requests, the specifications recommended by URSA has only 1 core and/or 2GB memory more than the optimal ones. Meanwhile, we measure and show the performance of each workload with the resource specification recommended by URSA and the optimal specification in Figure~\ref{fig:solo_up}. The figure shows that all the workloads achieve the required performance improvement with the specifications recommended by URSA. URSA can recommend the appropriate resource configuration for a workload because URSA can accurately predict the scaling surface of the workload based on system indexes and hardware event statistics. Based on the precise scaling surface, URSA can easily identify the smallest specification that satisfies the required performance.

{\it URSA can recommend the ``just-enough'' resource specification that achieves the workloads' performance target.}


\subsection{Scenario 2: Reducing Rent Cost}
\label{sec:case 2}

In this experiment, to emulate the scenario that tenants over-rent resources for ensuring high performance, we assume the origin resource specification of each workload in the validation set is (12C,16G) that consists of 12 cores and 16GB memory. 

\begin{figure}[tbp]
	\centering
	\includegraphics[width=0.8\columnwidth]{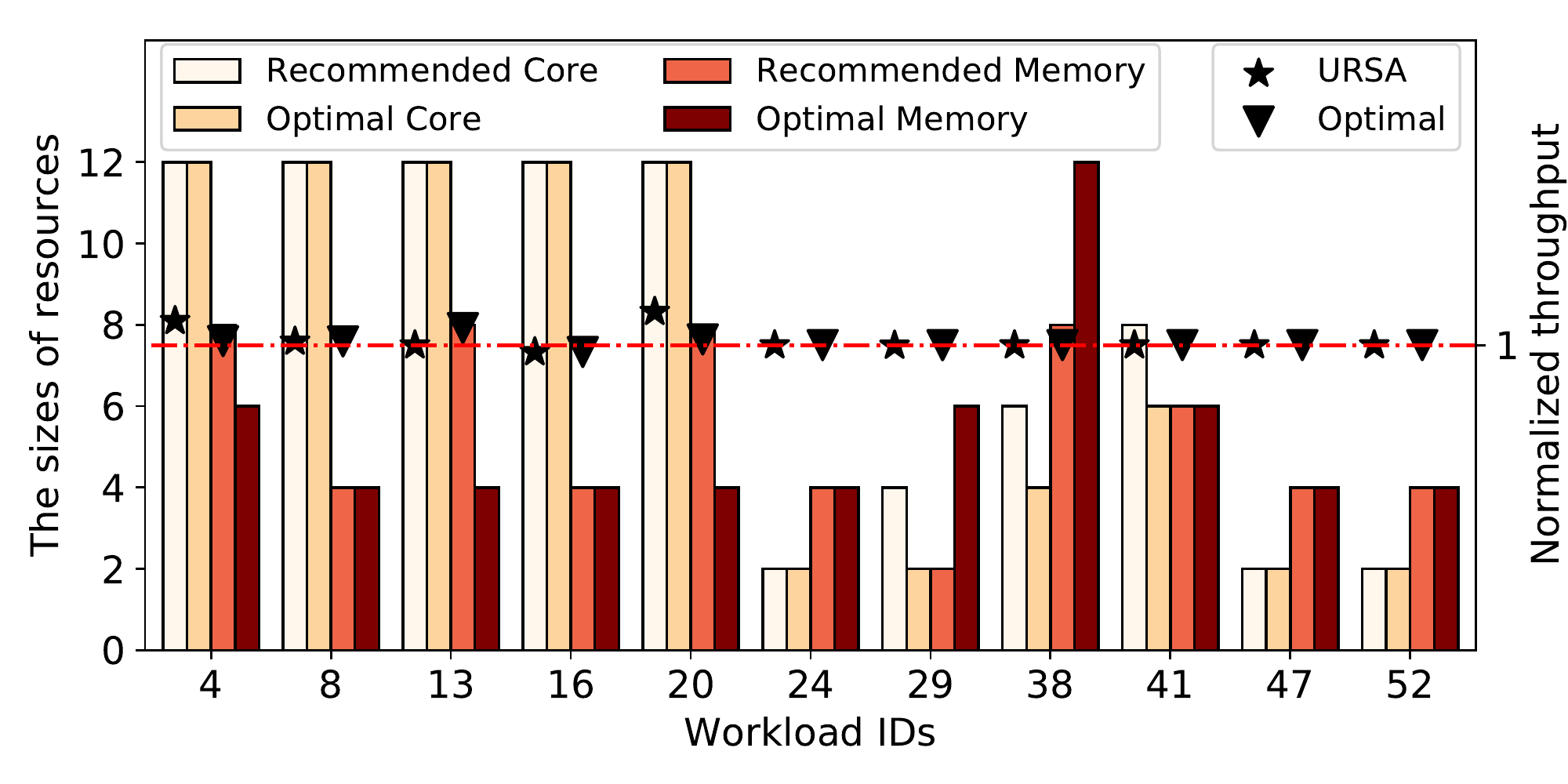}
	\vspace{-1mm}
	\caption{Specifications recommended by URSA and the optimal specifications for reducing rent cost.}
	\label{fig:solo_down}
	\vspace{-3mm}
\end{figure}

Figure~\ref{fig:solo_down} shows the specification recommended by URSA and the optimal (i.e., the smallest) specification identified by searching through all the possible resource specifications. If there are multiple local optimal specifications, the specification with the least number of cores is selected as the optimal configuration. Observed from the figure, URSA identifies the optimal resource specifications for 5 out of 11 requests. For the other requests, the specifications recommended by URSA only has 1 more core and/or 4GB memory. For the 11 workloads, the specifications recommended by URSA reduces 43.6\% cores and 65.5\% memory compared with the original specifications without degrading the original performance, and uses 7.7\% more cores and 3.4\% more memory than the optimal specifications. Meanwhile, Figure~\ref{fig:solo_down} also shows that the specifications recommended by URSA do not hurt the performance of all the workloads in the validation test.  

{\it URSA can accurately recommend the smallest specifications for database workloads without hurting the original performance.}

\subsection{Alleviating Shared Resource Contention}
In this experiment, we evaluate the effectiveness of the contention-aware scheduling engine in alleviating shared resource contention. As for the experimental setup, we co-locate 56 database workloads that are randomly selected from the validation set on our 7-node dbPaaS Cloud. We use 56 database workloads because the dbPaaS Cloud has overall $96\times 7 = 672$ cores and $256GB\times 7=1792GB$ memory that can support at most 56 workloads if they use the largest specification. To emulate real-system scenario, the origin specification of each database workload is configured to be ($m$C, $n$G), where $m$ and $n$ are randomly integers ($1\le m \le 12$, $2 \le n \le 16$).

We compare URSA with the \emph{LeastRequestedPriority} scheduling strategy in Kubernetes~\cite{k8s}.
When adopting URSA to schedule the workloads, URSA recommends a smaller specification for each database workload without hurting its performance, and schedules them using the contention-aware scheduling engine. If LeastRequestedPriority is adopted, for each node, the scheduling engine calculates the percentage of its free resources would be used if the workload is scheduled to it. The node with the smallest percentage is selected to host the workload.
The LeastRequestedPriority strategy balances the workloads based on their required resources, and the resources used on each node are basically balanced. 
For the fairness of comparison, we run the above experiments 10 times and report the result of each test.  


We first report the effectiveness of URSA in reducing resource consumption without hurting the system-wide performance. To quantify the system-wide performance, we first calculate the slowdown of each workload at co-location. Equation~\ref{eq:slowdown} calculates the \emph{slowdown} of a workload (denoted by $SD$) when it is co-located with other workloads due to shared resource contention. In the equation, $T_{colo}$ and $T_{solo}$ represent the throughput of the workload at co-location and when it monopolizes a node, respectively.
\begin{equation}
\vspace{-1mm}
\label{eq:slowdown}
SD=T_{colo}/T_{solo}
\vspace{-1mm}
\end{equation}

Based on Equation~\ref{eq:slowdown}, Equation~\ref{eq:throughput} calculates the aggregated performance of the workloads where $SD_i$ is the slowdown of the $i$-th workload and $N_w$ is the number of database workloads. The reason we do not use the aggregated TPS to be the system-wide performance is that different workloads run difference transactions.
\begin{equation}
\label{eq:throughput}
\vspace{-1mm}
P_{sys}=\sum_{i=1}^{N_{w}} SD_{i}
\end{equation}

\begin{figure}
	\centering
	\subfloat[\emph{Resources usage}]{
		\includegraphics[width=0.4\columnwidth]{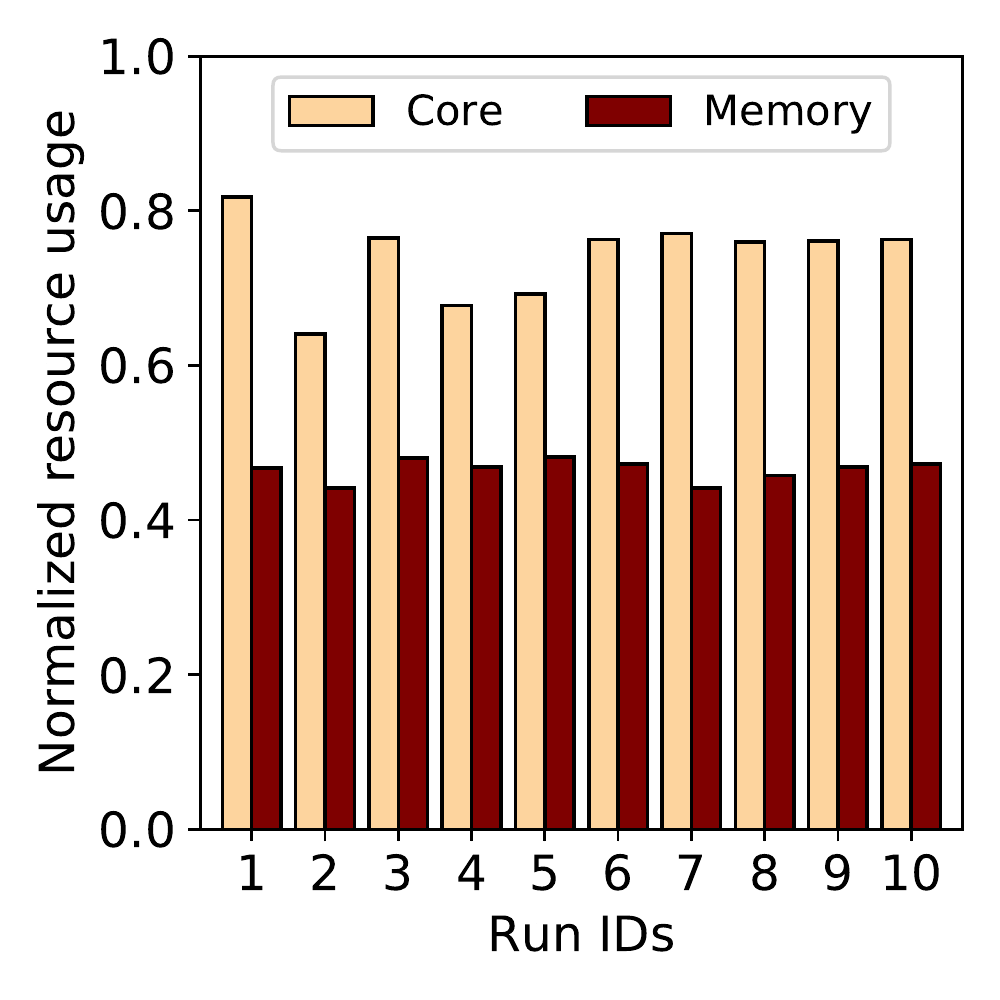}
		\label{fig:colo_results}
	}
	\subfloat[\emph{Throughput and unfairness}]{
		\includegraphics[width=0.4\columnwidth]{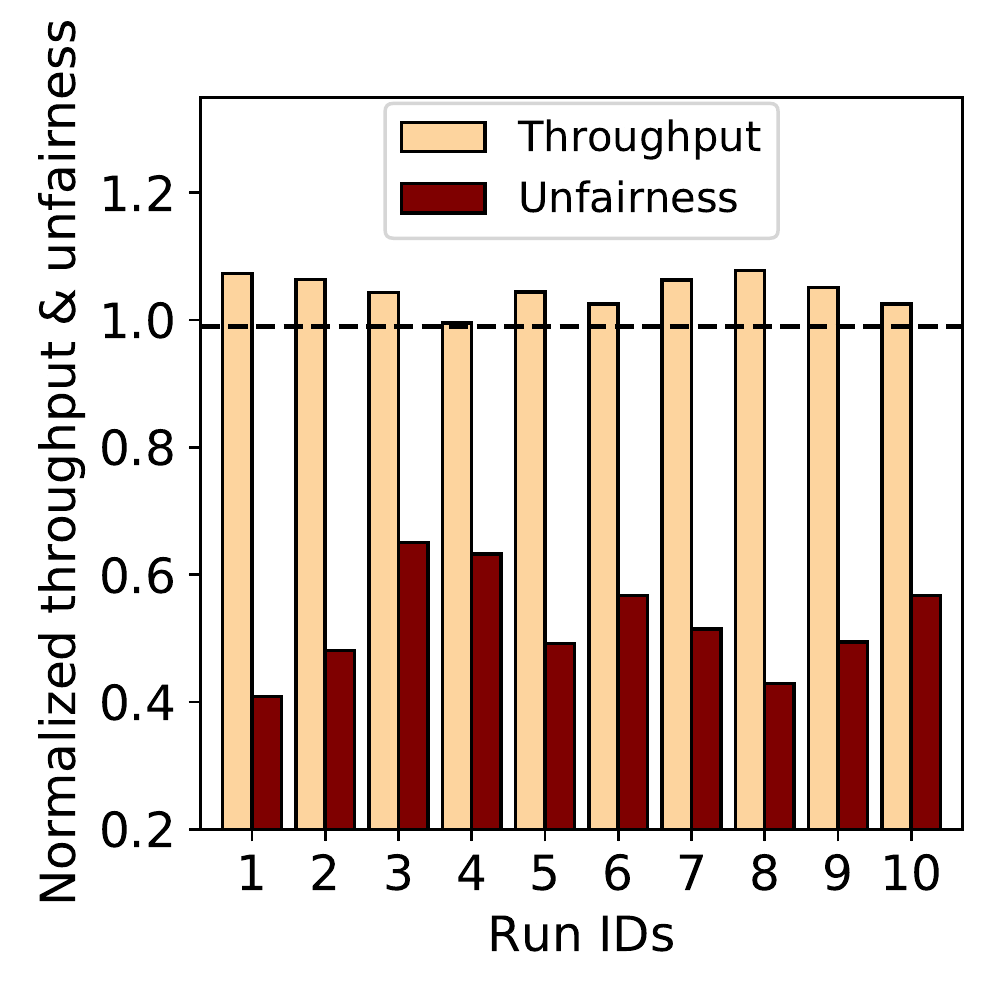}
		\label{fig:colo-throughput}
	}
	\caption{Resource usage, throughput and unfairness of URSA in the co-location experiment.}
	\vspace{-4mm}
\end{figure}

Figure~\ref{fig:colo_results} shows the overall number of cores and the size of memory used by the 56 workloads with the specifications recommended by URSA normalized to their counterparts in the origin specifications. Observed from the figure, URSA reduces the number of cores and memory space used by the 56 workloads in all the 10 tests. On average, URSA reduces 25.9\% of cores, 53.4\% of memory space. Figure~\ref{fig:colo-throughput} shows the aggregated performance of the workloads when they are scheduled with URSA normalized to their performance with the LeastRequestedPriority strategy. URSA does not hurt the aggregated performance in all the 10 tests.

We also compare the performance fairness of the workloads at co-location with URSA and the LeastRequestedPriority strategy. Equation~\ref{eq:unfairness} calculates the unfairness of the scheduling, where $N_{w}$ is the number of workloads deployed in the Cloud and $SD_i$ is the slowdown of the \emph{i}-th workload calculated in Equation~\ref{eq:slowdown}. 
\begin{equation}
\label{eq:unfairness}
unfaireness=\frac{\max \limits_{1<\forall i<N_{W}} SD_{i}- \min \limits_{1<\forall i<N_{W}} SD_{i}}{\max \limits_{1<\forall i<N_{W}} SD_{i}}
\end{equation}

Figure~\ref{fig:colo-throughput} shows the performance fairness of the database workloads with URSA normalized to the fairness with LeastRequestedPriority strategy. Observed from the figure, URSA greatly reduces the performance unfairness between database workloads in the Cloud. On average, URSA reduces 47.6\% of the unfairness. The improved performance fairness originates from the alleviating of serious shared resource contention on some nodes. 

{\it URSA reduces the resource usage of database workloads without hurting the system-wide performance. It also improves the performance fairness between the co-located workloads.}

\subsection{Analyzing URSA Overhead}
\label{sec:Overhead}
In this section, we present the overhead of URSA. The offline overhead of the model training in URSA is acceptable. We collect the performance data of 55 database workloads on a single node in 3 days. This process can be shortened by collecting the data on multiple machines in parallel. For dbPaaS providers, they have much more various database workloads which can generate good coverage of the scaling characteristic space to produce a well-trained model. For example, if a dbPaaS provider uses 20 nodes to collect training data from 2000 database workloads, the time cost of collecting the training data is around 6 days which is acceptable. The trained model is valid unless the Cloud provider replaces current nodes with new generation machines.

As for the overhead of the online predicting, operating systems provide low overhead tools, such as perf and cgroups, to collect hardware counter statistics and system indexes. The time of collecting the indexes for a database is shorter than 1s, and the scaling surface predicting completes in 30$\mu$s. 

The overhead of the interference estimator is positively related to the numbers of levels of pressure and sensitivity. The more levels are, the more time needed by the interference estimator. In the experiment in section~\ref{sec:case 2}, the pressure on the shared resources and the sensitivity to shared resource contention have 20 levels, and the estimator quantifies the pressure and sensitivity of a workload in 2 minutes that is affordable for a long-running database workload.

The time needed to search for an appropriate node for a database workload is correlated with the number of nodes in the Cloud. Our measurement shows that the scheduling engine finds an appropriate node for a database workload on our 7-node  Cloud in 0.12ms. And, it is possible to parallelize the search procedure if the number of nodes is large to further reduce the overhead.

\section{Conclusion}
\label{sec:conclusion}
URSA predicts the scaling surface of a database workload based on hardware counter statistics and system-level indexes. Based on the predicted scaling surface, URSA recommends an appropriate specification to meet the performance requirements. URSA can accurately predicts the scaling surface of database workloads with the prediction error smaller than 5.8\%. In addition, URSA quantifies the pressure on shared resource and sensitivity to shared resource contention for each workload and alleviates serious shared resource contention when scheduling. Our experimental results show that URSA reduces 25.9\% of CPU usage and 53.4\% of memory usage for database workloads while satisfying their performance requirements. Meanwhile, URSA reduces the performance unfairness by 47.6\% compared with the LeastRequestedPriority policy.

\bibliographystyle{IEEEtran}
\bibliography{ref}

\end{document}